\documentclass[%
 reprint,
 amsmath,amssymb,
 aps,pra
]{revtex4-2}

\usepackage[table]{xcolor}

\usepackage{siunitx}
\usepackage{graphicx}
\usepackage{dcolumn}
\usepackage{bm}
\usepackage[unicode=true, colorlinks=true, citecolor={blue!80!black}, urlcolor={blue!50!black}, linkcolor = {blue!80!black}]{hyperref}
\usepackage{easyReview}
\usepackage{adjustbox}
\usepackage{xcolor}
\usepackage[T1]{fontenc}

\begin{document}

\preprint{QuTech/AndersenLab/cQEDteam}

\title{Gate-tunable kinetic inductance parametric amplifier}

\author{Lukas Johannes Splitthoff$^{1,2}$}
\email{l.j.splitthoff@gmail.com}
\author{Jaap Joachim Wesdorp$^{1,2}$}
\author{Marta Pita-Vidal$^{1,2}$}
\author{Arno Bargerbos$^{1,2}$}
\author{Yu Liu$^{3}$}
\author{Christian Kraglund Andersen$^{1,2}$}

\affiliation{$^1$QuTech, Delft University of Technology, Delft 2628 CJ, The Netherlands}
\affiliation{$^2$Kavli Institute for Nanoscience, Delft University of Technology, Delft 2628 CJ, The Netherlands}
\affiliation{$^3$Center for Quantum Devices, Niels Bohr Institute, University of Copenhagen, 2100 Copenhagen, Denmark}

\date{\today}

\begin{abstract}
    Superconducting parametric amplifiers play a crucial role in the preparation and readout of quantum states at microwave frequencies, enabling high-fidelity measurements of superconducting qubits. Most existing implementations of these amplifiers rely on the nonlinearity from Josephson junctions, superconducting quantum interference devices or disordered superconductors. Additionally, frequency tunability arises typically from either flux or current biasing. In contrast, semiconductor-based parametric amplifiers are tunable by local electric fields, which impose a smaller thermal load on the cryogenic setup than current and flux biasing and lead to vanishing crosstalk to other on-chip quantum systems. In this work, we present a gate-tunable parametric amplifier that operates without Josephson junctions, utilizing a proximitized semiconducting nanowire. This design achieves near-quantum-limited performance, featuring more than 20 dB gain and a 30 MHz gain-bandwidth product. The absence of Josephson junctions allows for advantages, including substantial saturation powers of \SI{-120}{dBm}, magnetic field compatibility up to \SI{500}{\milli T} and frequency tunability over a range of \SI{15}{MHz}. Our realization of a parametric amplifier supplements efforts towards gate-controlled superconducting electronics, further advancing the abilities for high-performing quantum measurements of semiconductor-based and superconducting quantum devices.
\end{abstract}

\maketitle

\section{Introduction}
In the pursuit of advancing quantum technologies, the extraction and amplification of weak quantum signals have emerged as crucial challenges in all known qubit platforms, especially those operating in cryogenic environments. The here required amplification chains for weak signals must be carefully designed to exhibit specific characteristics, including high gain, wide amplification bandwidth, and large saturation power while maintaining minimal added noise, to maximize the signal to noise ratio for high-performing readout~\cite{Eichler2014, Roy2016, Aumentado2020}.
Superconducting parametric amplifiers based on Josephson junctions already play a pivotal role at the first stage of amplification chains in superconducting qubit platforms. By adding only the minimal amount of noise permitted by the laws of quantum mechanics, these amplifiers have proven invaluable for amplifying weak microwave signals encoding the quantum state of superconducting qubits~\cite{Yurke1989, Clerk2010, Abdo2011, Vijay2011, Krantz2016, Walter2017}.
Other quantum systems, including spin qubit implementations~\cite{Mi2018,Borjans2020,Kobayashi2021,Philips2022,Mills2022}, and novel types of hybrid semiconductor-superconductor qubits~\cite{deLange2015, Pita-Vidal2023, Pita-Vidal2023_2, Dvir2023}, as well as condensed matter experiments~\cite{Wang2019, Portolés2022} are generally operated under magnetic fields and therefore require magnetic field compatible parametric amplifiers. An additionally desired feature is minimal crosstalk to sensitive structures in the vicinity of the amplifier, such as flux-tunable qubits. However, conventional superconducting parametric amplifiers with flux-~\cite{Yamamoto2008, Castellanos2007, Castellanos2008,Zhong2013, Zhou2014} or current-biased~\cite{Parker2022} control are impractical to operate near flux-sensitive elements. Surpassing these limitations necessitates the development of novel types of parametric amplifiers combining magnetic field compatibility with a new source of tunability. 

Recent technological advances in the integration of exotic hetero-structures into superconducting circuits, such as hybrid superconducting-semiconducting nano-structures~\cite{deLange2015, Larsen2015, Casparis2018, Kroll2019, Splitthoff2022, Phan2022-1, Strickland2023}, graphene Josephson junctions~\cite{Kroll2018, Schmidt2018, Wang2019}, or carbon-nanotube junctions~\cite{Mergenthaler2021} have enabled the realization of electrostatic control of supercurrents in superconducting circuits. This additional method of tunability has already opened up new avenues to build novel types of parametric amplifiers from graphene~\cite{Butseraen2022, Sarkar2022} and proximitized semiconductors~\cite{Phan2022} further advancing the development of densely packed superconducting electronics due to minimal crosstalk.
Concurrently, magnetic field compatibility of parametric amplifiers has been achieved through the use of superconducting thin films acting as kinetic inductance material~\cite{Khalifa2023, Xu2023, Vine2023}. 

In this work, we leverage the continuous superconducting aluminium thin film on an InAs nanowire, which can be seen as a chain of infinitesimally short, gate tunable Josephson junctions to experimentally demonstrate a magnetic-field-compatible, gate-tunable kinetic inductance parametric amplifier composed of an InAs/Al nanowire shunting a NbTiN coplanar waveguide resonator~\cite{Splitthoff2022}. Notably, the parametric amplifier presented in this work features a gate-tunable amplification window, and magnetic field compatibility resulting from the superconducting thin film and the Josephson junction free design. Moreover, the amplifier exhibits a substantial saturation power and increased resilience against electrostatic discharge compared to Josephson junction based implementations, owing to the continuous superconducting film.

\section{Experimental setup}
\begin{figure}
    \centering
    \includegraphics{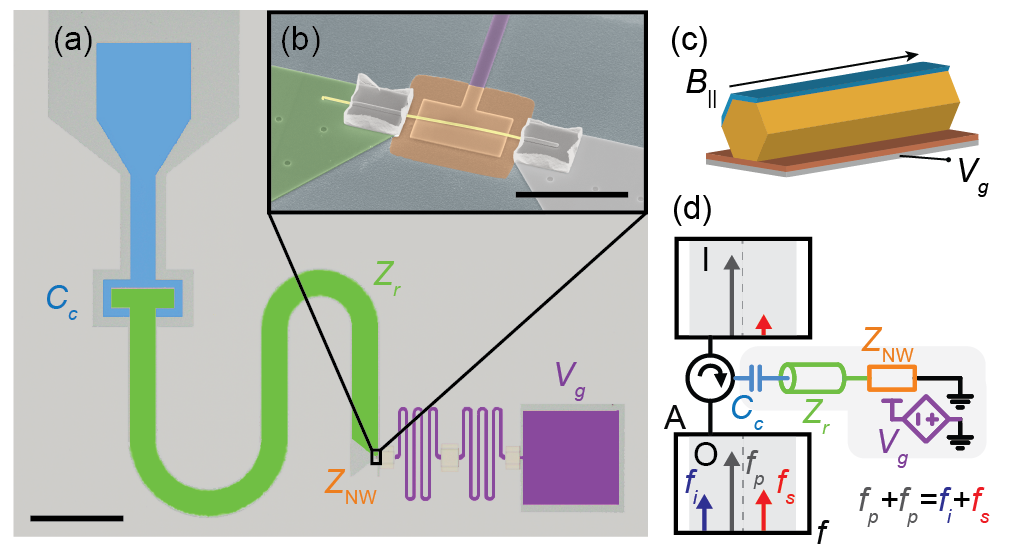}
    \caption{Device and equivalent circuit. (a) False-colored optical microscope image of a gate-tunable parametric amplifier comprised of a quarter-wave coplanar waveguide resonator (green) that is capacitively coupled to the reflection port (blue) and shorted to ground (light grey) via a proximitized nanowire (yellow). The gate line with a 5th order Chebyshev LC filter (purple) suppresses the cross talk between the resonator and gate line. (scale bar $\SI{250}{\micro \meter}$). (b) Electron micrograph of the nanowire segment. The nanowire is connected with thick NbTiN patches to the resonator and ground. (scale bar $\SI{5}{\micro \meter}$). (c) Schematic representation of the InAs nanowire (yellow) which is proximitized with a continuous Al shell (blue) on 2 facets and positioned on a voltage gate (grey) with a dielectric (orange). (d) Equivalent circuit of the parametric amplifier and representation of the 4-wave mixing operation showing the signal amplitude $A$ as a function of frequency $f$. At the input (I), a pump tone (grey) at a frequency $f_p$ and a signal (red) with a frequency $f_s$ enter the nonlinear resonator. At the output (O), the signal is amplified and there is an additional idler tone at a frequency $f_i$. The dashed line indicates the central frequency of the undriven resonator and the grey box is the amplifier bandwidth.}
    \label{fig:Fig1}
\end{figure}

In our realization, we use a quarter-wave coplanar waveguide resonator that is capacitively coupled to a launch pad at one end and shorted to ground via a proximitized InAs/Al nanowire, see Fig.~\ref{fig:Fig1}a. The frequency of the resonator is controlled using a DC-voltage applied to the gate line (see purple structure in Fig.~\ref{fig:Fig1}a). To minimize losses of the coplanar waveguide, we implement a 5th order Chebyshev LC filter which suppresses coupling to the gate line by at least \SI{50}{dB} within a frequency range of $4-\SI{8}{GHz}$, see Appendix~\ref{app:device} for additional details on the device design and the LC filter. For electrostatic control, we follow Ref.~\cite{Splitthoff2022} and place the $l=\SI{4.5}{\micro \meter}$ long nanowire on an electrostatic voltage gate that extends for $\SI{3.5}{\micro \meter}$ below the nanowire segment, see Fig.~\ref{fig:Fig1}b. The nanowire section is directly connected to the central conductor of the resonator and to ground via thick NbTiN patches. The schematic representation of the proximitized nanowire in Fig.~\ref{fig:Fig1}c highlights the continuous Al shell covering two facets of the InAs nanowire~\cite{Krogstrup2015}. We maximize the magnetic field compatibility by aligning the magnetic field $B_\parallel$ parallel with the nanowire. 

The nanowire-shunted resonator is expected to exhibit a nonlinear Kerr-type behavior when driven near resonance, where the resonance frequency shifts linearly with respect to the number of microwave photons occupying the resonator mode, see also Appendix~\ref{app:ext_data}. As a result, it is natural to utilize the nonlinear resonator as a parametric amplifier in the non-degenerate mode, which preserves the phase of the input signal. This mode of operation takes advantage of the 4-wave mixing process occurring within a Kerr-oscillator. This process involves the interaction of two pump photons, at frequency $f_p$, and one signal photon, at frequency $f_s$, leading to the conversion of a pair of pump photons into an idler photon at $f_i$ and an additional signal photon, which therefore leads to amplification of the signal tone (as illustrated in Fig.~\ref{fig:Fig1}d). We integrate a circulator at the input to the device to ensure that the signal and pump photons pass through the parametric amplifier, while also maintaining the desired directionality from the input to the output spectrum. Overall, this experimental configuration enables efficient phase preserving amplification of the input signal, see also Appendix~\ref{app:setup} for a detailed description of the experimental setup.

\section{Experimental Results}

\subsection{Gate-tunable nonlinear resonator}
\begin{figure}
    \centering
    \includegraphics{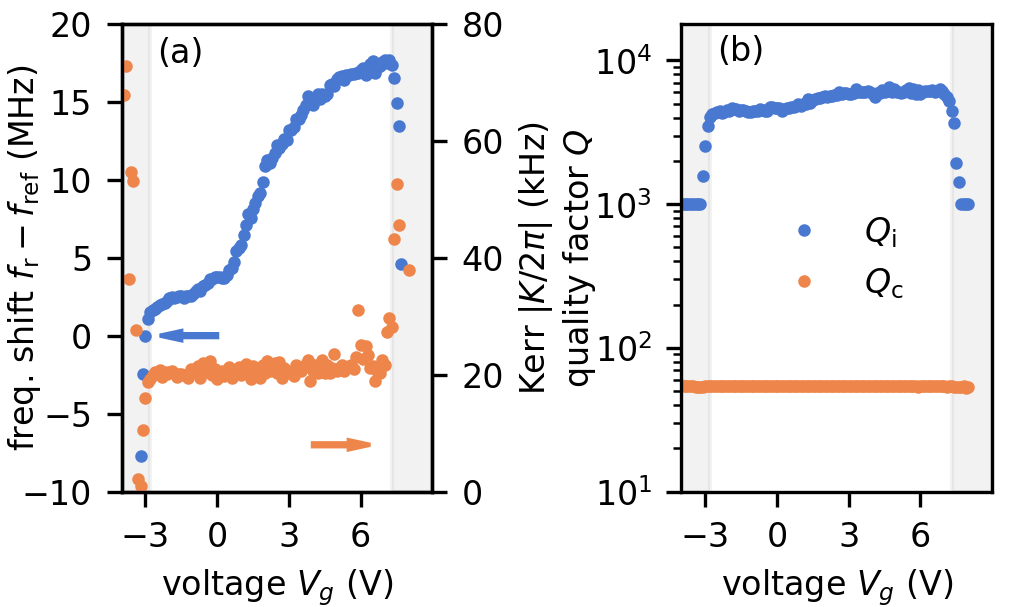}
    \caption{Gate-tunable nonlinear resonator. (a) Frequency shift (blue, left axis) relative to the reference frequency $f_{\text{ref}}$ measured at $V_g=\SI{-3}{V}$ and Kerr coefficient $K$ (orange, right axis) as function of gate voltage, $V_g$. (b) Internal quality factor $Q_i$ (blue) and coupling quality factor $Q_c$ (orange) as a function of gate voltage $V_g$. The grey shaded areas indicate regions of suppressed resonator quality for gate voltages below $\SI{-3}{V}$ and above $\SI{7}{V}$. }
    \label{fig:Fig2}
\end{figure}

To characterize the parametric amplifier, we start by measuring the complex transmission parameter $S_{21}$ between input and output port to extract the resonance frequency of the resonator as well as the internal and coupling quality factors. These quantities are extracted from a fit to a complex Lorentzian as expected for linear reflection type resonators~\cite{Khalil2012, Probst2015}. 
The resonance frequency is tunable with gate voltage and we observe a monotonic increase in resonance frequency for voltages within the range of $\SI{-3}{V}$ to $\SI{7}{V}$, see Fig.~\ref{fig:Fig2}a. The shift of about $\SI{15}{M \hertz}$ arises from a change in the kinetic inductance associated with the proximitized nanowire, as shown in Ref.~\cite{Splitthoff2022}. The nearby voltage gate controls the charge carrier density in the nanowire and hence the normal state conductivity, which dominates the high frequency response. At the same time, the superconducting gap closes only partially with increasing gate voltage. 
To characterize the nonlinearity of the resonator, we increase the power of the spectroscopy tone and we extract a Kerr-coefficient $K$ as the linear slope of the resonator frequency as a function of inferred photon number inside the resonator. We find $K\approx\SI{20}{\kilo \hertz}$ for all gate voltages, which ranges in between typical values reported for parametric amplifiers based on a single junction~\cite{Butseraen2022} and for kinetic inductance based amplifiers~\cite{Khalifa2023}. The flat response versus gate voltage facilitates a robust tune up of the device in contrast to previous voltage-tunable junction based systems, where the Kerr coefficient strongly depends on the gate voltage~\cite{Butseraen2022, Sarkar2022, Phan2022}.
The internal quality factor increases slightly over the accessible gate voltage range starting at $Q_i \sim \SI{4000}{}$, whereas the coupling quality factor stays nearly constant at $Q_c \sim \SI{50}{}$. The increase in $Q_i$ could be attributed to the finite gap at all gate voltages and the addition of further conduction channels with more positive gate voltage, although a detailed study of the internal loss mechanism is beyond what can be extracted from simple spectroscopy measurements. In addition, we observe a small, although negligible for the applications considered here, reduction of $Q_i$ due to nonlinear losses (see Fig.~\ref{fig:FigSKerrcoefficient}).
Outside of the $\SI{-3}{V}$ to $\SI{7}{V}$ voltage range, indicated by grey boxes in Fig.~\ref{fig:Fig2}, the resonator coherence is suppressed by gate leakage. The n-doped Si substrate forms a Schottky junction below about $\SI{-3}{V}$ negative voltage bias leading to the injection of quasi-particles. Above $\SI{7}{V}$ we observe the breakdown of the SiN gate dielectric. These range limitations could be mitigated in the future by alternative material choices for the dielectric media.

\subsection{Amplifier characteristics}
\begin{figure}
    \centering
    \includegraphics{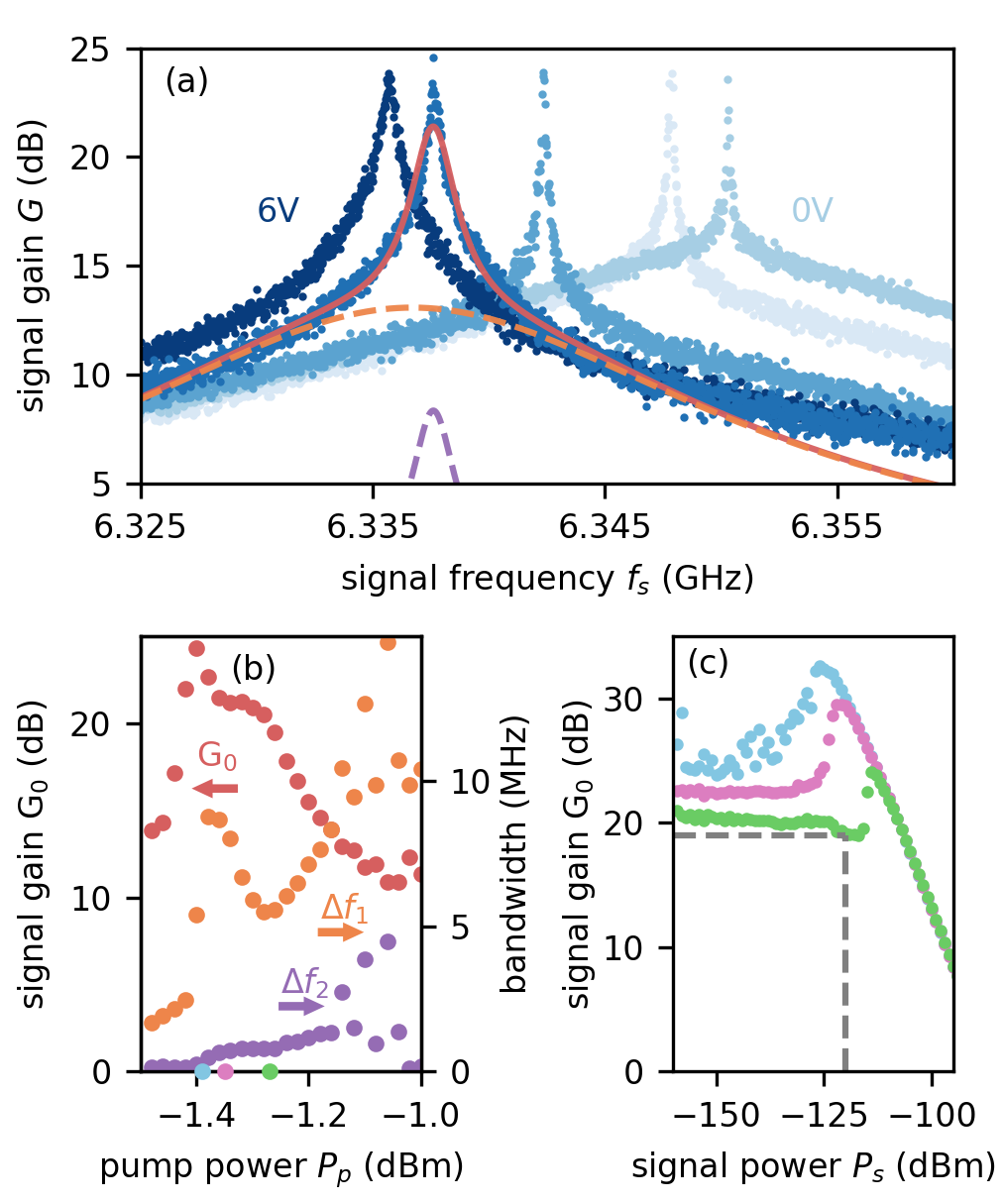}
    \caption{Amplifier characteristics. 
    (a) Signal gain versus signal frequency for the different gate voltage settings [\SI{6}{V}, \SI{4}{V}, \SI{2}{V}, \SI{0}{V}, \SI{-2}{V}] from dark to light shades. The solid (red) curve shows the best fit of a double Lorentzian to the gain curve. The dashed (purple and orange) curves show the individual single Lorentzian curves composing the fits. 
    (b) Signal gain, $G_0$ and bandwidths of the two Lorentzian curves, $\Delta f_1$ and $\Delta f_2$, as a function of pump power at the signal generator, $P_p$, at room temperature.
    (c) Signal gain as a function of referred signal power at the PA input, $P_s$, at the amplifier input for three different pump powers indicated on the x axis of (b) with color-matching markers.}
    \label{fig:Fig3}
\end{figure}

Next, we demonstrate that the nonlinear resonator can be operated as a parametric amplifier. We apply a strong pump tone slightly detuned from the bare resonance frequency and we measure signal gain for signal frequencies in the vicinity of the pump, see Fig.~\ref{fig:Fig3}(a). Notably, we find that by changing the gate voltage, the points of maximal gain shift to a different frequencies, exemplified for five different gate voltages in Fig.~\ref{fig:Fig3}(a).
Specifically, we find that the amplifier exhibits more than $\SI{20}{dB}$ gain in all tested configurations yielding a gate tunable amplification window of over $\SI{15}{MHz}$. In nominally identical devices but different experimental realizations, we have observed accessible frequency tuning ranges of more than \SI{70}{MHz}. It is important to note that the gate dependence of the parametric amplifier's working point exhibits a non-monotonic behavior, which is in contrast to the monotonic frequency response observed in Fig.~\ref{fig:Fig2}a, resulting from an interplay between the linear and nonlinear inductances. However, while the gate dependence of the optimal working point cannot be a priori predicted, we find only a weak hysteresis versus gate voltage, as reported in earlier work~\cite{Splitthoff2022}, which allows for a robust and efficient tune-up as well as stable operation over days of the parametric amplifier.

We fit the signal gain curve with a double Lorentzian model which captures both the up to \SI{15}{dB} broadband amplification window (orange) and the narrowband mode (purple) centered around the pump frequency, see Fig.~\ref{fig:Fig3}a. This double Lorentzian behavior may arise due to the presence of a parasitic mode, most likely a box mode stemming from the sample enclosure. Note that the fit underestimates the maximal gain observed in the gain curve.
Despite this complex mode structure, we extract the maximal signal gain $G_0$ of the fitted response versus pump power $P_p$ and find a stable working range of \SI{0.1}{dB} with more than \SI{20}{dB} gain, see Fig.~\ref{fig:Fig3}b which exemplifies the trend at a gate setting of \SI{+4}{V}. 
The bandwidths of the two respective modes $\Delta f_1$ and $\Delta f_2$ evolve non-monotonically as a function of power. While $\Delta f_1$ increases for pump powers above $\SI{-1.25}{dB}$, $\Delta f_2$ stays nearly constant. 
The trend of the gain curve (red) and contribution to the bandwidth of the broad mode (orange) can be explained by considering a single mode Kerr-resonator model for a fixed pump frequency $\omega_p$ and a variable pump strength $\xi$. Since the optimal pump frequency depends on the pump power, the system deviates from the optimal pump condition and the low order approximation yielding a constant gain-bandwidth product breaks down. Instead we expect a minimal bandwidth (orange curve) at the point of maximal gain (red), which is in agreement with the data shown in Fig. \ref{fig:Fig3}b around \SI{-1.25}{dBm}. For larger pump powers, the bandwidth increases and the gain decreases as expected but due to the second mode we do not reach the coupling limited gain-bandwidth product. This simple description does not capture the regime of smaller pump powers below \SI{-1.4}{dBm} when the distinction of the two modes becomes unreliable due to the small gain and small bandwidth. Consequently, the gain-bandwidth product of neither mode reaches the bandwidth limit of \SI{128}{MHz} set by the coupling quality factor of the resonator, as would be expected for an ideal single-mode Kerr-resonator close to the optimal pump condition.  
To characterize the dynamic range of amplification, we record the maximal signal gain $G_0$, directly extracted from the data, versus signal power, as shown in Fig.~\ref{fig:Fig3}c for three different pump powers as indicated with colored dots in Fig.~\ref{fig:Fig3}b. With increasing signal power the signal gain overall reduces due to a deviation from the stiff pumping condition \cite{Eichler2014, Roy2016}. 
For a weak signal gain of \SI{20}{dB} we observe a \SI{1}{dB} compression point of \SI{-120}{dBm}, which is on par with the saturation powers reported in work on other capacitively coupled SQUID, SQUID array, single junction and gate tunable junction implementation \cite{Butseraen2022, Sarkar2022, Phan2022} (see also Tab. \ref{tab:amplifieroverview}) and should allow for frequency-multiplexed qubit readout given a sufficiently large amplification bandwidth.
The sharp gain rise in the signal gain before its overall decrease, sometimes referred to as ``shark fin'', results from an interplay between the input signal and the internal dynamics of the parametric amplifier due to higher order nonlinear terms at negative pump detuning. This gain rise phenomenon has been experimentally observed~\cite{Zhou2014, Narla2014} and theoretically described~\cite{Liu2017, Sivak2019, Liu2020}. 

\subsection{Noise performance}
\begin{figure}
    \centering
    \includegraphics{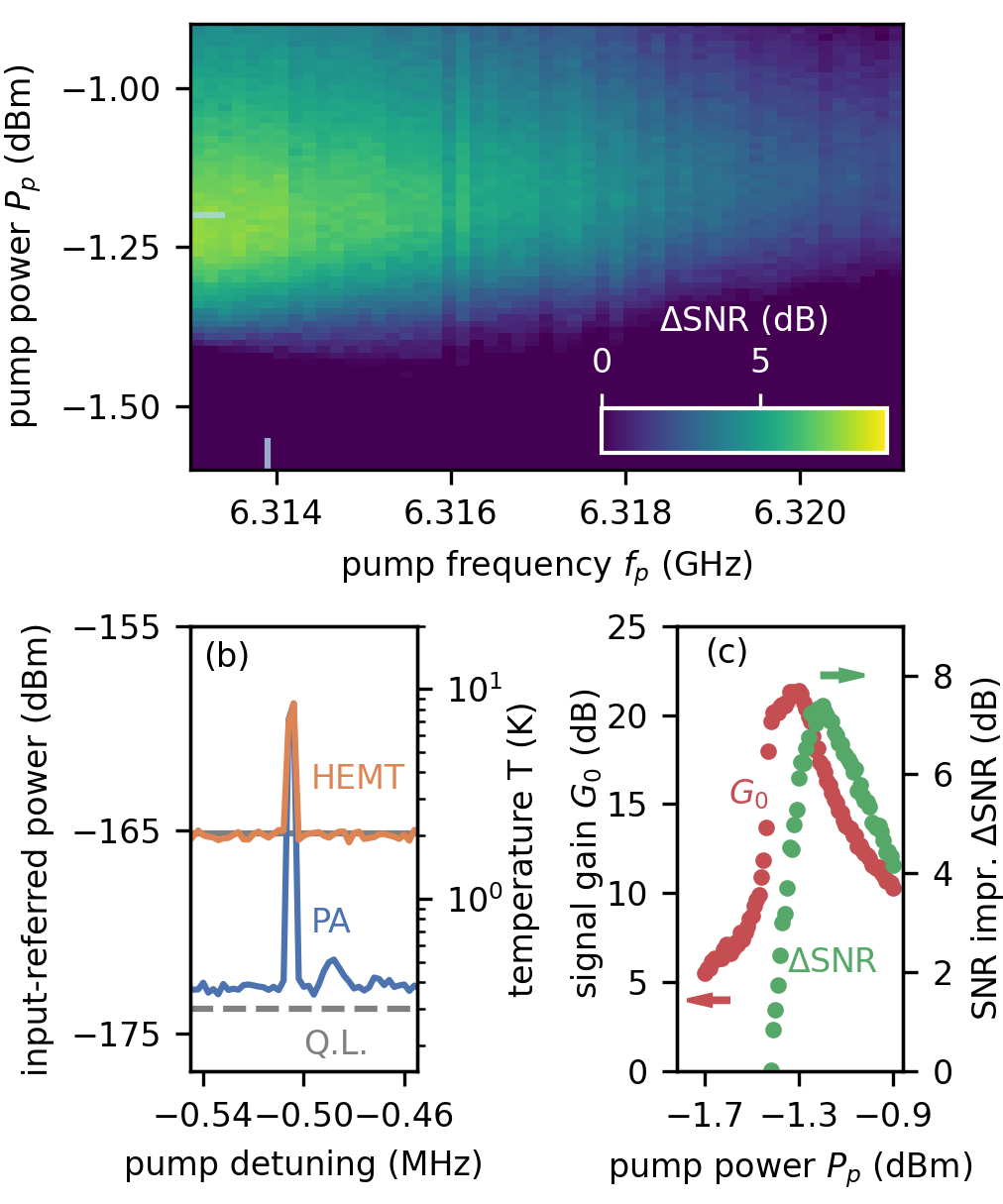}
    \caption{Signal to noise (SNR) improvement. (a) Change in the signal to noise ratio as a function of pump power, $P_p$, and the pump frequency, $f_p$, for a fixed detuning between signal and pump frequency of \SI{0.5}{MHz}. Light blue lines indicate the working point used for the trace shown in (b). (b) Input-referred power spectrum with signal tone at \SI{6.3134}{GHz} and pump tone at \SI{6.3139}{GHz} in the undriven (orange) and driven (blue) case. The dashed lines indicate the quantum limit (Q.L.) and the specified noise temperature of the HEMT for a resolution bandwidth of \SI{1}{kHz}. For comparison, the equivalent temperature is provided on the right axis.
    (c) Signal gain (red, left axis) and SNR improvement (green, right axis) versus pump power for $f_p=\SI{6.3139}{GHz}$.}
    \label{fig:Fig4 Noise}
\end{figure}

While large gain is desirable for parametric amplifiers, it should come without the cost of additional added noise beyond the quantum limit, in order to not degrade the amplifier performance. We generally quantify the quality of a signal by considering the signal to noise ratio (SNR). Specifically, we compare the power level of the signal to the power level when the signal is turned off or, more practically, the power level slightly detuned from the signal. For an amplifier, we can define the SNR improvement as the difference of the SNR between the on and off state of the amplifier $\Delta \text{SNR} = \text{SNR}_{\text{on}} - \text{SNR}_{\text{off}}$ which serves as a good measure for its noise characteristic. For the parametric amplifier that we study here, we find that the SNR improvement depends strongly on the pump power and pump frequency as shown in
Fig.~\ref{fig:Fig4 Noise}a, measured at a gate voltage of \SI{+4}{V} and at a signal-pump detuning of $\Delta=\SI{0.5}{MHz}$. The SNR improvement reaches a maximum of about \SI{7}{dB}, which is about \SI{2}{dB} lower than the theoretical maximum set by the difference between the noise temperature of the HEMT and the quantum limit (QL)~\cite{Caves1982}, see Fig.~\ref{fig:Fig4 Noise}(b). It is worth noticing that the maximal SNR improvement versus pump power presented in Fig.~\ref{fig:Fig4 Noise}c does not coincide with the maximum in the gain curve, but shows a peak at about \SI{15}{dB} signal gain. Note that we slightly compress an amplifier at room temperature when approaching maximal gain, which might have led to the slight shift between the point of maximal gain with respect to the point of maximal SNR improvement in Fig.~\ref{fig:Fig4 Noise}c. Also note that the working point of the parametric amplifier has shifted by about \SI{10}{MHz} compared to the data shown in Fig.~\ref{fig:Fig2}a after many gate scans, which might have led to a different charge configuration in the nanowire~\cite{Meyer2023} and hence to a slightly different working point. 
In Fig.~\ref{fig:Fig4 Noise}b, we compare the spectrum obtained for a single pump tone between the driven and undriven state of the parametric amplifier. In the undriven case, the noise is limited by the high-mobility-electron transistor (HEMT) amplifier at the 4K stage of the cryostat. The referred power, and hence the effective noise temperature, is estimated based on a line calibration supported by an estimate based on the dispersive shift of an adjacent on-chip transmon qubit connected to the input port (see Appendix~\ref{sec:transmon}). In the undriven case we observe the signal tone at \SI{6.3134}{GHz} and the pump tone leaking through the output of the signal generator, while the noise floor is set by the noise temperature of the HEMT, which is close to the manufacture-specified value of \SI{2.2}{K} indicated as grey horizontal line. In the driven case the signal amplitude increases together with the appearance of an idler tone of about the same magnitude, while the noise floor drops below the noise floor of the HEMT, showcasing the usefulness of the here presented parametric amplifier. While the improvement in SNR is noteworthy, we also observe that the performance is not fully quantum limited as we do not reach the quantum limit of \SI{303}{\milli K}, indicated with a dashed grey line. The inability to reach the quantum limit signals the presences of an additional loss channel in the system, which is consistent with the observations of an additional mode in Fig.~\ref{fig:Fig3}.

\subsection{Magnetic field compatibility}
\begin{figure}
    \centering
    \includegraphics{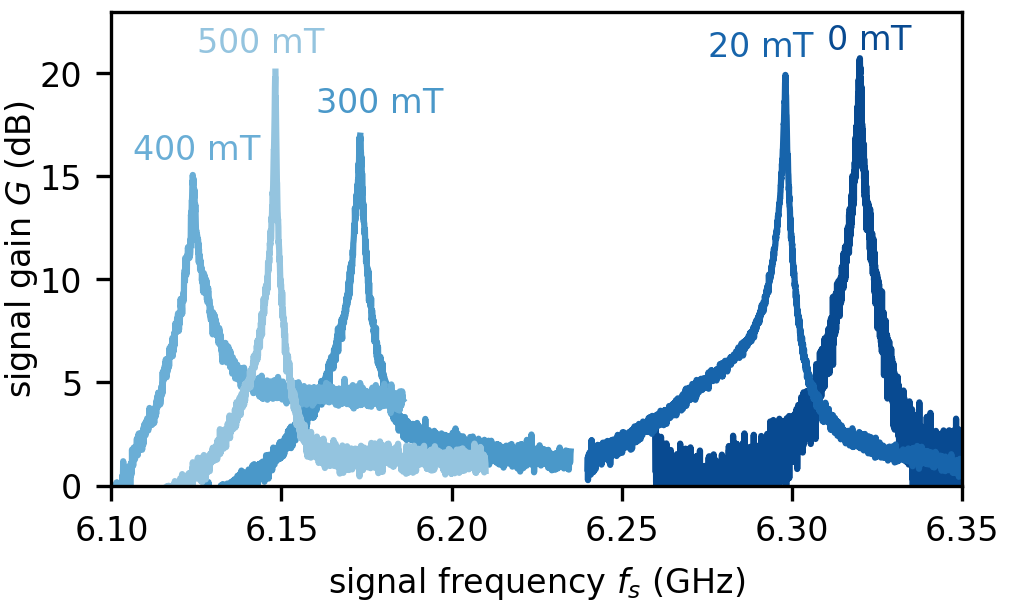}
    \caption{Magnetic field compatibility. Signal gain versus signal frequency for the different parallel magnetic field strength \SI{0}{\milli T}, \SI{20}{\milli T}, \SI{300}{\milli T}, \SI{400}{\milli T}, and \SI{500}{\milli T}.}
    \label{fig:Fig5 field}
\end{figure}

The continuous superconducting thin film renders amplification at high magnetic field strength possible, since the superconducting gap is not locally suppressed inside a Josephson junction potentially leading to interference effects. In particular, we are interested in magnetic fields of up to a few hundred millitesla as used for a variety of spin~\cite{Philips2022} and hybrid superconductor-semiconductor devices~\cite{Dvir2023, Pita-Vidal2023}. When the magnetic field, $B_{\parallel}$, is aligned with the nanowire direction, the compatibility with field should be maximal and we observe up to \SI{20}{dB} gain at fields up to \SI{0.5}{T}, see Fig.~\ref{fig:Fig5 field}. Moreover, at \SI{0.5}{T}, we see maximal gain at a \SI{170}{MHz} lower signal frequency compared to the zero field setting. The overall shift of the amplification window towards smaller frequencies along with the increase in the Kerr coefficients follows from the continuous suppression of the superconducting gap with magnetic field~\cite{Splitthoff2022}. The non-monotonicity of the frequency shift and the lower maximal gain can be attributed to the nucleation of vortices in the hetero-structure of finite thickness and to small field misalignments~\cite{Krause2022}; a scenario which is most likely unavoidable in real applications of this amplifier. The distortion of the gain curves in Fig.\ref{fig:Fig5 field} arises most likely from the presence of the broad box mode by Fano interference \cite{Rieger2023}. We did not explore the magnetic field dependence of the parametric amplifier beyond \SI{0.5}{T} neither did we characterise the noise performance at elevated magnetic fields due to technical limitations imposed by the dilution refrigerator which prevented us from maintaining an elevated field while keeping the base temperature cold. However, we do not expect a significant change in the noise performance as the resonator loss rate $\gamma$ remains smaller than the coupling strength $\kappa$. While an increase in the resonator loss rate $\gamma$ is unavoidable, previous work on similar hybrid systems allowed the realization of a transmon qubit operated at \SI{1}{T} \cite{Kringhoj2021}, which shows that the fabrication of coherent devices at elevated magnetic field is possible. Note that this data set was taken in a consequent cool down, which again led to a different charge configuration in the nanowire and hence a different working point.   

\section{Conclusion}
We experimentally demonstrated a prototype of a gate-tunable kinetic inductance parametric amplifier. This system features a gate tunable amplification window of up to \SI{15}{MHz} due to the hybrid superconducting-semiconducting nano-structure, magnetic field compatibility up to at least \SI{500}{mT} due to the superconducting thin film, a sizeable saturation power of \SI{-120}{dBm} and minimal electrostatic discharge sensitivity due to the continuous superconducting film. 
The here reported tuning range of the amplification window should already be sufficient to render the frequency matching with other superconducting resonators made from low kinetic inductance film and hence low fabrication induced frequency variability \cite{Andersen2020_2} possible.
Furthermore, we expect that the local gate voltage control, as demonstrated here, will exhibit minimal crosstalk to other voltage controlled element due to the absence of finite and long range supercurrents induced by the DC control. Thus, this ability to tune the nonlinear resonator might render large scale implementations of voltage-controlled superconducting electronics with dense packing possible. 

While this first demonstration of a gate tunable kinetic inductance parametric amplifier achieves the main objectives of magnetic field compatibility, gate tunability and a useful saturation power, further improvements must be made without additionally compromising any of these feature. Therefore, further work should be devoted to the improvement of the dielectric environment to widen the tuning range without being limited by current leakage at higher voltages. Moreover, the optimization of the inductance ratio of the proximitized nanowire to the resonator is required to further enhance the saturation power of the amplifier. In this work, we only study 4-wave mixing operations, but 3-wave-mixing should be accessible by driving the nonlinear resonator via an oscillating voltage gate connected to the proximitized nanowire or alternatively a DC current bias through the nonlinear inductor. We expect that 3-wave-mixing allows for a large spectral separation of the pump tone and the signal tone, alleviating frequency crowding in the amplification window and saturation of higher amplification stages. 

\section*{Acknowledgements} 
We thank Peter Krogstrup for the nanowire growth. We also thank Alessandro Miano, Patrick Winkel, Nicolas Zapata González and Ioan Pop for valuable discussions. 
This research was co-funded by the allowance for Top consortia for Knowledge and Innovation (TKI) from the Dutch Ministry of Economic Affairs and the Microsoft Quantum initiative. 
CKA acknowledges financial support from the Dutch Research Council (NWO).

\section*{Author contributions}
LJS, and CKA conceived the experiment. LJS fabricated the devices, and acquired and analysed the data with help from JJW, MPV and AB. YL provided the proximitized nanowires. LJS and CKA wrote the manuscript with input from all other co-authors. CKA supervised the project. 

\section*{Data availability}
The raw data and the analysis script underlying all figures in this manuscript  are available online \cite{Splitthoff2023}.

\appendix
\counterwithin{figure}{section}

\section{Design and fabrication}
\label{app:device}

\subsection{Design}
\begin{figure}
    \centering
    \includegraphics{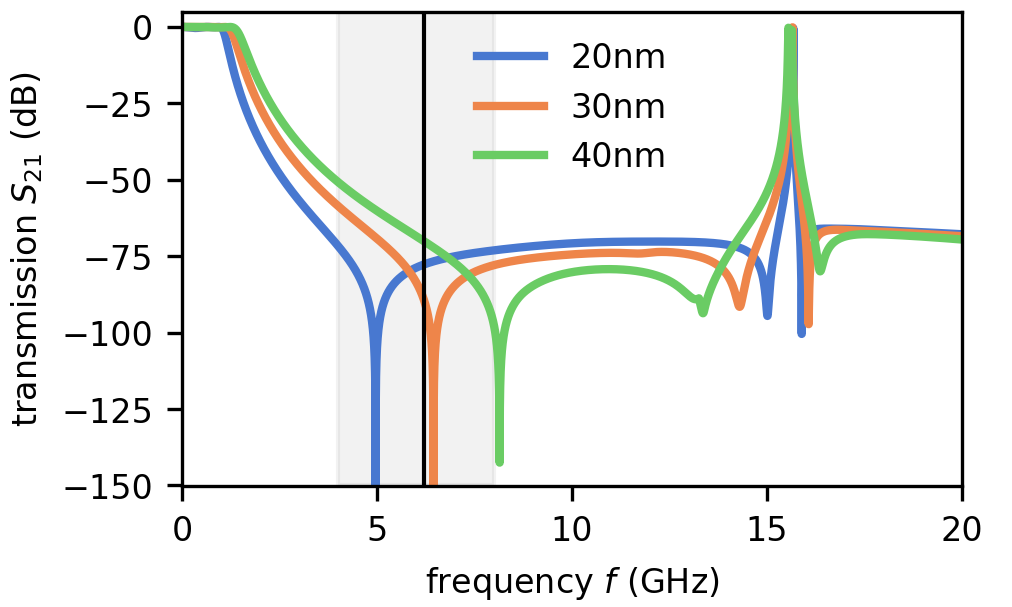}
    \caption{Filter response: Simulated transmission parameter $S_{21}$ versus frequency of a 5th order Chebyshev filter for three different dielectric layer thicknesses. The grey box indicates the 4-\SI{8}{GHz} range. The black vertical line indicates the resonance frequency of the undriven amplifier.}
    \label{fig:FigS Filter}
\end{figure}

We design the circuitry using a customised design framework \cite{Gehring2019}. 
The loaded quarter-wave resonator is designed to have a total capacitance of $C_0=\SI{511}{fF}$, a total inductance of $L_0=\SI{1.07}{nH}$ for the unloaded transmission line resonator, a capacitance to the input port of $C_c=\SI{55}{fF}$ and an inductance of $L_{NW}=\SI{0.79}{nH}$ for the proximitized nanowire. We note that if the nanowire segment was a single Josephson junction, we would expect a critical current of $I_c = \frac{\Phi_0}{2\pi L_{\text{NW}}}\approx \SI{0.42}{\mu A}$ for this inductance. Instead in DC experiments on nanowires from the same batch, we measured a critical current of around $3-\SI{17}{\mu A}$, which quantitatively supports the description of the nanowire as kinetic inductance element.  
The nanowire inductance fraction in this configuration is $\alpha = \frac{L_{NW}}{L_{NW}+L_0} = 0.42$. The quarter-wave transmission line design frequency separates best the fundamental mode from higher harmonics simulated to be above \SI{19.2}{GHz}.
To suppress cross talk between the amplifier mode and the voltage gate, we add a 5th order Chebyshev low pass filter in between. The filter is composed of three parallel plate capacitors and two meandering inductors as shown in Fig. \ref{fig:Fig1}a. As expected for a low pass filter, we find that the filter suppresses high frequency transmission, see Fig. \ref{fig:FigS Filter} which shows the low pass filter response from finite element simulations for three different thicknesses of the dielectric material SiN separating the two superconducting plates. The filter therefore suppresses the coupling between amplifier and gate by more than \SI{50}{dB} in the 4-\SI{8}{GHz} range (grey box) also in the case of small fabrication imperfections potentially resulting in the SiN film thickness. The filter structure also exhibits self-resonances due to its finite length at around \SI{15.5}{GHz}. In microwave simulations, we do not find any additional or common mode due to the filter structure around the resonance frequency of the undriven amplifier that could explain the appearance of an additional mode in the driven case.

\subsection{Fabrication}
We fabricate the resonator circuit and the gate lines from a \SI{40}{\nano \meter}-thick sputtered NbTiN film (kinetic inductance \SI{4}{\pico \henry \, \Box^{-1}}) on high resistivity n-doped Si. We pattern the NbTiN film using e-beam lithography and SF$_6$/O$_2$ based reactive ion etching. \SI{30}{\nano \meter}-thick plasma enhanced chemical vapour deposition (PECVD) SiN defined by a buffered oxide etch serves as bottom gate dielectric. We transfer the two-facet InAs/Al nanowire on top of the SiN bottom gate using a nano-manipulator. The InAs nanowires were grown by vapor–liquid–solid (VLS) growth with a diameter of \SI{110(5)}{\nano \meter}, and nominal thickness of the Al of \SI{6}{\nano \meter} \cite{Krogstrup2015}. We electrically contact the nanowires to the circuit via lift-off defined \SI{150}{\nano \meter}-thick sputtered NbTiN leads after prior Ar milling to minimize the contact resistance. 

\subsection{Reproducibility}
For this project, we have prepared 13 samples with 4 amplifiers each of which we have measured 7 samples with in total 12 parametric amplifiers in 12 cooldowns, of which 6 amplifiers exhibited more than 20dB gain. The remaining devices were discarded because of fabrication imperfections or excluded because of high room temperature resistances of the nanowire segment due to high nanowire to NbTiN contact resistance.

\section{Kerr coefficient}

\begin{figure}
    \centering
    \includegraphics{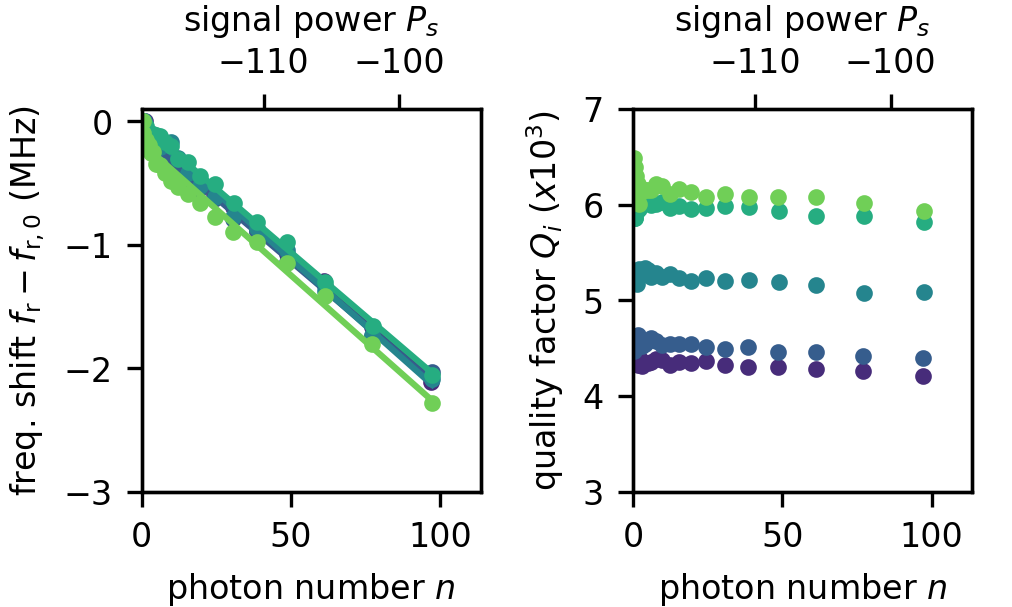}
    \caption{Nonlinear resonator: (a) Frequency shift and (b) internal quality factor for five different gate voltages between \SI{-2.5}{V} and \SI{5.5}{V} versus intra-cavity photon number.}
    \label{fig:FigSKerrcoefficient}
\end{figure}

\label{app:ext_data}

The parametric amplifier is a nonlinear resonator where the nonlinearity, as explained in the main text, manifests as a frequency shift proportional to the number of photons in the resonators. We extract this proportionality constant, the Kerr coefficient $K$, from a measurement of the frequency shift versus intra-resonator photon number, see
Fig. \ref{fig:FigSKerrcoefficient} which shows this measurement for different gate voltages. The frequency shifts are displayed here with respect to the resonator frequency measured at the weakest signal power $f_r,0$. We observe a similar slope for all gate voltage between \SI{-2.5}{V} and \SI{5.5}{V} which results in a Kerr coefficient of about \SI{20}{kHz}. 
As shown in Fig. \ref{fig:FigSKerrcoefficient}, the internal quality factor increases with gate voltage from $Q_{i,\SI{-2.5}{V}}=\SI{4363}{}$ to $Q_{i,\SI{5.5}{V}}=\SI{6483}{}$ and for all gate voltages, we see a weak dependence of $Q_i$ as a function of the intra-resonator photon number.  

\section{Measurement setup}
\label{app:setup}

\begin{figure*}
    \centering
    \includegraphics{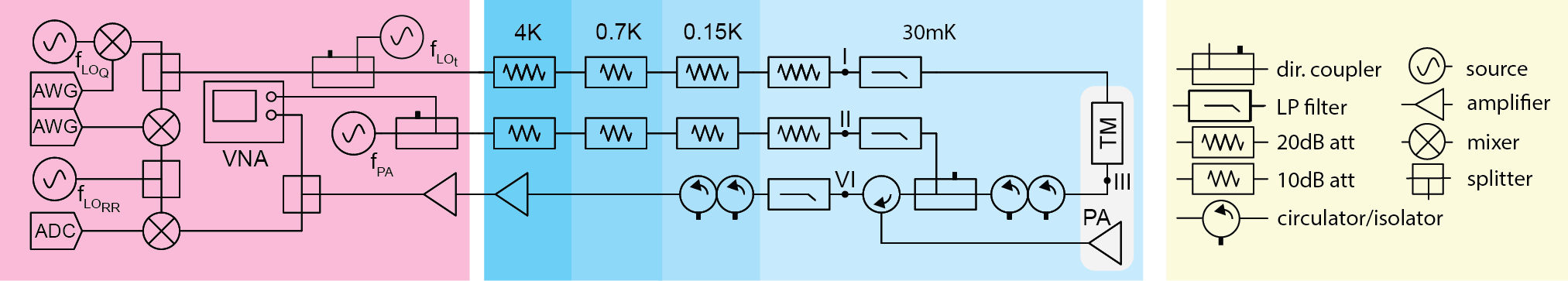}
    \caption{Experimental setup. 
    (red) Room temperature control and readout electronics: A signal generator, $f_{PA}$ and a vector network analyzer (VNA) are combined with a directional coupler and connected to the drive line of the parametric amplifier. Two arbitrary waveform generators (AWG) provide modulated waveforms which are upconverted to microwave frequencies with the local oscillators set by two signal generators, $f_{LO_Q}$ and $f_{LO_{RR}}$, for the transmon qubit drive and the readout. The output signal is split into a signal path to the vector network analyer and towards the analog to digital convertor (ADC). 
    (blue) Microwave components installed in the dilution refrigerator, including two input and one output line. The sample (grey) hosts the readout resonator, the transmon qubit and the parametric amplifier. Two strongly attenuated and low pass filtered input lines serve as readout resonator and transmon qubit drive line and as parametric amplifier drive line. The output line is amplified using one cryogenic high-electron mobility transistor (HEMT) and one room temperature HEMT.}
    \label{fig: wiring diagram}
\end{figure*}

The 6x6 mm chip hosting the parametric amplifier, and also a readout resonator and transmon qubit (see Sec. \ref{sec:transmon}), is glued with GE varnish onto a gold plated copper mount and electrically connected to a printed circuit board using Al wire-bonds. 

The room temperature control and readout electronics (red box in Fig.~\ref{fig: wiring diagram}) control the sample under test. A signal generator and a vector network analyzer are connected to the drive line for the parametric amplifier. Amplitude modulated waveforms provided by the two other signal generators and two baseband arbitrary waveform generators drive the transmon qubit and the readout resonator. 
The output signal is split into a signal path to the vector network analyer and towards the amplitude demodulation unit. 
Microwave components installed in the dilution refrigerator (blue areas in Fig.~\ref{fig: wiring diagram}) include two strongly attenuated and low pass filtered input lines and one output line, and the sample hosting the readout resonator, the transmon qubit and the parametric amplifier. The output line is amplified using one HEMT operating at cryogenic temperatures (\SI{4}{K}) and one HEMT operating at room temperature as well as the parametric amplifier under test. 

\begin{figure}
   \centering
   \includegraphics{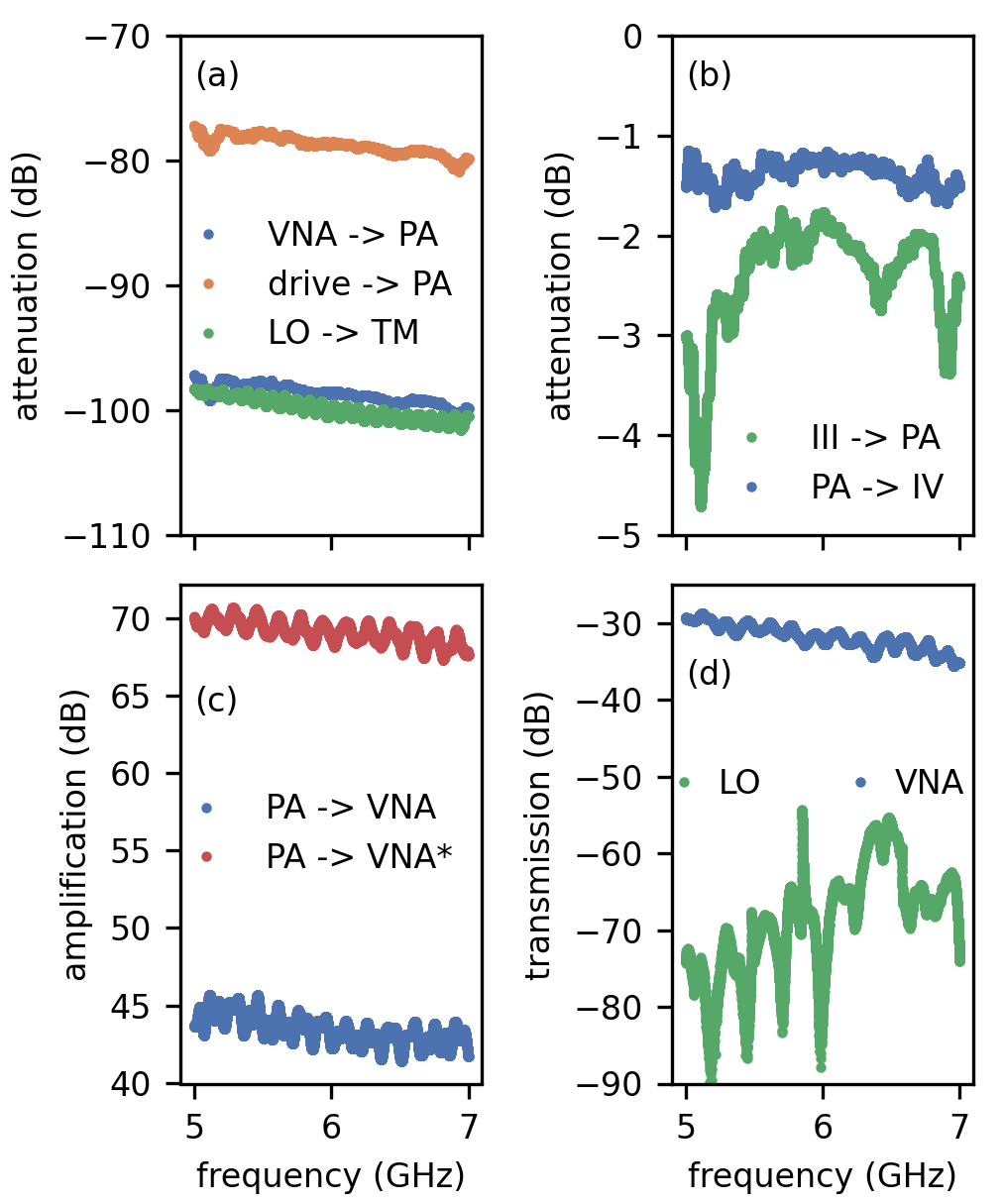}
   \caption{Transmission calibration of measurement chain and individual components measured with a VNA. (a) Input line transmission at room temperature from VNA to PA input (blue), from PA drive signal generator to PA input (orange) and from the test tone signal generator to the TM input (green). (b) Interconnection transmission at room temperature from TM output (III in Fig. \ref{fig: wiring diagram}) to PA input and from PA output to HEMT input (IV in Fig. \ref{fig: wiring diagram}). (c) Net amplification of the readout chain between PA output and VNA input for two different configurations. (d) Transmission measurement in the cold state with PA off from VNA to VNA port (blue) and LO test signal generator to VNA input (green)}
   \label{fig:FigC2}
\end{figure}

The estimation of the line attenuation and amplification between room temperature electronics and the parametric amplifier is based on a combination of transmission measurements at cold temperature and room temperature and summarized in Fig.~\ref{fig:FigC2}.  
The probe line between VNA and parametric amplifier has \SI{-99}{dB} attenuation at \SI{6.4}{GHz}, which is composed of \SI{50}{dB} cold attenuation between the VNA output and PA input, \SI{40}{dB} from the two directional couplers, \SI{3.3}{dB} attenuation from the eccosorb filters at \SI{6.4}{GHz}, \SI{3.2}{dB} cable and connector loss at room temperature, and the insertion loss of low pass filter, circulator and copper cables connecting to the PA of about \SI{2.5}{dB}, see the blue line in Fig.~\ref{fig:FigC2}a . The connection between the PA drive signal generator and the parametric amplifier has \SI{20}{dB} less attenuation due to the through connection through the first directional coupler at room temperature. The attenuation between the LO test signal generator (LO), that generates the test tone for the SNR improvement measurement, and the input to the transmon readout resonator (TM) is \SI{-100}{dB}. 
Fig.~\ref{fig:FigC2}b shows the attenuation of the interconnections before and after the parametric amplifier, in particular the connection from the transmon readout resonator output (III) to the parametric amplifier input (PA) with \SI{-2.5}{dB} attenuation and the connection between PA output and HEMT input (IV) with \SI{-1.3}{dB} attenuation. 
Fig.~\ref{fig:FigC2}c shows the net amplification of the readout chain for two different tested configurations. The data presented in this manuscript is based on the "PA $\rightarrow$ VNA$^*$" configuration with \SI{70.7}{dB} gain provided by the cryo-HEMT (LNF LNC4\_8C s/n 844H, $P_{1dB,out, cryo}=\SI{-8}{dBm}$) and the LNF room temperature HEMT (LNF-LNR4\_8F\_ART, $P_{1dB,out, rt}=\SI{0}{dBm}$), which we tested component-wise at room temperature.
Finally, we can compare the component-wise characterization for the input lines (Fig.~\ref{fig:FigC2}a) and output line (Fig.~\ref{fig:FigC2}c) at room temperature with the transmission measurement through the entire setup in the cold state, presented in Fig.~\ref{fig:FigC2}d.
The sum of the individual input and output line characterizations agrees up to \SI{4}{dB} with the total transmission measurement from VNA to VNA transmission as shown in the wiring diagram in Fig.~\ref{fig: wiring diagram}. This discrepancy might arise from multiple reconnections between the components and a different thermal state of the components in the respective measurements.

This lines calibration suggests that the typically used pump power level at the PA of $P_p=\SI{-1.2}{dBm}-\SI{78.9}{dB}=\SI{-80.1}{dBm}$ does not compress the cyro-HEMT with input saturation power $P_{1dB,in, cryo}\approx\SI{-48}{dBm}$. However, the pump power level after the cryo-amplifier is about $P_p\approx\SI{-40}{dBm}$ which is on the level as the \SI{1}{dB} input compression power of the room temperature HEMT of $P_{1dB,in, rt}\approx\SI{-42}{dBm}$. Consequently, the pump tone together with the amplified signal tone may saturate the room temperature amplifier in this configuration of the measurement setup, which may distort the gain curves and also cause the shift between the point of maximal gain and the point of maximal SNR improvement, observed in Fig.~\ref{fig:Fig4 Noise}c. To mitigate the saturation of the room temperature amplifier, we considered an alternative configuration of the readout chain including a \SI{20}{dB} attenuator before an alternative room temperature amplifier with lower gain, but also lower \SI{1}{dB} compression point. The net amplification of this readout chain is shown as blue curve in Fig.~\ref{fig:FigC2}c. While the gain curves remained the same, we noticed that the readout chain is not limited by the HEMT noise temperature anymore but rather by the noise temperature of the spectrum analyzer, which would have complicated the interpretation of the noise characterization considerably.

\section{Cavity referred power using a transmon qubit coupled to a readout resonator} \label{sec:transmon}

\begin{figure}
    \centering
    \includegraphics{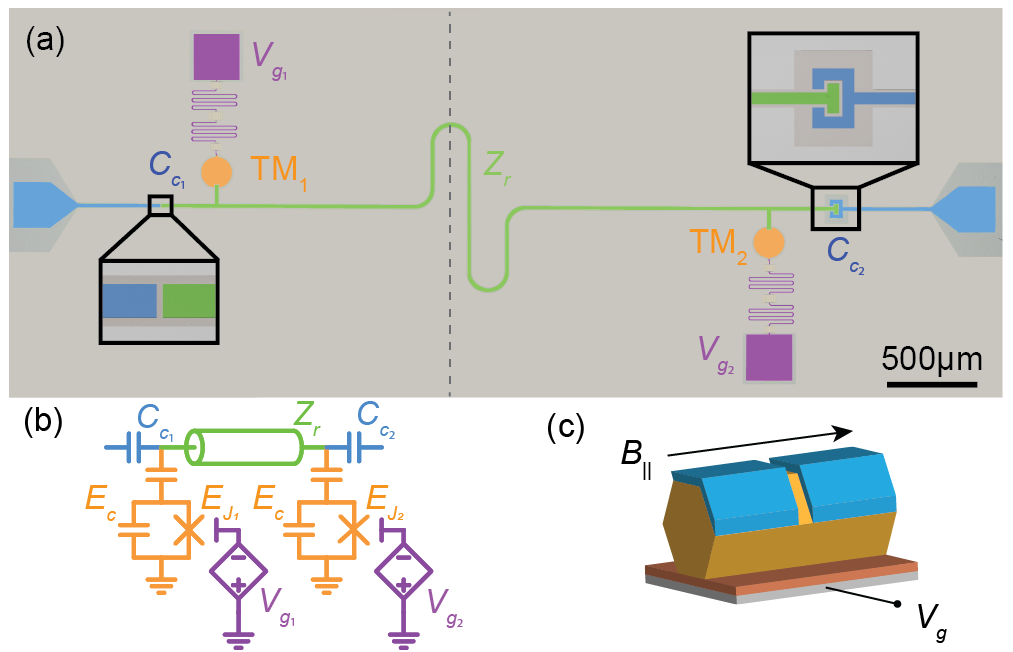}
    \caption{Transmon qubit and readout resonator. 
    (a) False-coloured optical microscope image of a readout resonator in transmission line configuration (green) that is capacitively connected to a weakly coupled input port (blue, left) and a strongly coupled output port (blue, right). Two single-island transmon qubits (orange) are capacitively coupled around the current nodes of readout resonator. The nanowire Josephson junctions of the transmon qubits are electrostatically controlled via low-pass filtered voltage gates (purple). The optical image has been cropped at the dashed grey line for better visibility, which reduces the displayed length of the resonator.
    (b) The equivalent circuit diagram of the device shown in (a). 
    (c) Schematic representation of the nanowire Josephson junction formed by an InAs nanowire which is partially proximitized with an Al shell on 2 facets and positioned on a voltage gate. }
    \label{fig: Transmon qubit layout}
\end{figure}

To further characterize the noise performance of the parametric amplifier and to showcase its usefulness in terms of co-operation with other quantum devices, we integrate two transmon qubits, which are dispersively coupled to a common readout resonator on the same device chip as the parametric amplifier. Fig.~\ref{fig: Transmon qubit layout}a shows a false-colored optical image of the transmission line readout resonator which is weakly coupled to the input port ($C_{C_1}$ small) and strongly coupled ($C_{C_2}$ large) to the output port. Two single-island transmon qubits~\cite{Barends2013} are capacitively coupled close to the current nodes of the resonator, see Fig. \ref{fig: Transmon qubit layout}b for the equivalent circuit diagram. The Josephson junctions of the two transmon qubits are formed by electrostatically controlled proximitized nanowires with etched segment as displayed in the schematic representation in Fig.~\ref{fig: Transmon qubit layout}c. The gate voltage controls the critical current of the Josephson junction, hence the Josephson energy $E_J$; but also a global magnetic field affects the Josephson energy. This tuning knob allows us to tune the transmon frequency $f_q$ in-situ as $h f_q (V_g) \approx \sqrt{8E_c E_J(V_g)}-E_c$ under the assumption of many low-transparency channels in the nanowire junction~\cite{Kringhoj2018}. Here, $E_c$ is the charging energy of the transmon island. 

\begin{figure}[h]
    \centering
    \includegraphics{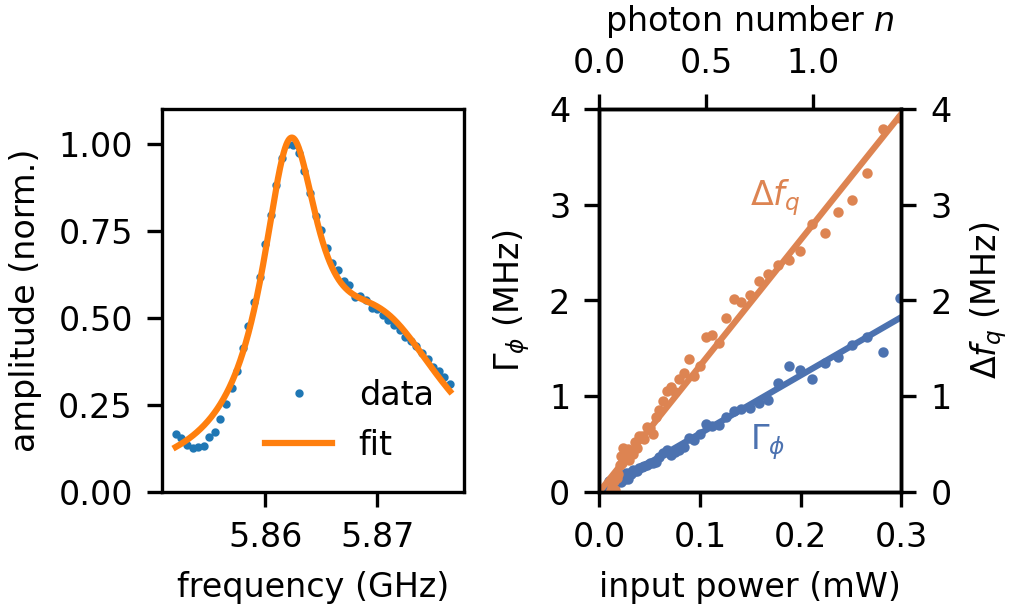}
    \caption{Estimation of intra-cavity photon number. (a) Transmission spectrum of readout resonator with double Lorentzian fit. (b) Measurement-induced dephasing (blue, left axis) and frequency shift (orange, right axis) as a function of variable input power of a readout tone with frequency $f_{LO_t}$ applied during a Ramsey experiment. The data is shown with linear fit to extract the slope. The equivalent intra-cavity photon number is indicated on the top axis.}
    \label{fig:FigSphotonnumber}
\end{figure}

To estimate the cavity referred power $P_{out}$, we independently measure the dispersive shift $\chi$ and the cavity coupling rate $\kappa$. We obtain $\chi$ from a variable strength Ramsey experiment \cite{Macklin2014, Bultink2018}, which allows us to extract the measurement-induced dephasing rate $\Gamma_\phi=\frac{8\chi^2n}{\kappa}$ and the qubit frequency shift $\Delta f_q=2\chi n$ as a function of a weak test tone at the readout resonator frequency applied using a signal generator at frequency $f_{LO_t}$ during the idling time of the qubit in the equal superposition state, see Fig. \ref{fig:FigSphotonnumber}b. 
From a linear fit to the dephasing rate $\Gamma_\phi$ versus input power at the signal generator output $P_{SG}$, we obtain the slope $\partial_P \Gamma_\phi$. Similarly, a linear fit to the frequency shift $\Delta f_q$ yields $\partial_P \Delta f_q$. Their ratio yields the dispersive shift $\chi$ expressed in the following equation:
\begin{equation}
    \chi = \frac{\kappa}{4} \frac{\partial_P \Gamma_\phi}{\partial_P \Delta f_q}
\end{equation}
From a linear fit to the frequency shift $\Delta f_q = 2\chi c P_{SG}$ we obtain the conversion factor $c$, which allows us to compute the intra-resonator photon number as $n=cP_{SG}$. 
Eventually, the cavity referred power at the resonator output is given by $P_{out}=\kappa h f_r n$. Due to the directional design of the transmission line resonator with $C_{C_2} \gg C_{C_1}$ and high internal quality factor $Q_i > Q_c$, we can approximate the coupling rate as $\kappa = \frac{f_r}{Q_{tot}}$, which is equivalent to the FWHM of the peak in the transmission measurement extracted from the double Lorentzian fit in Fig. \ref{fig:FigSphotonnumber}a.  
At one specific gate set point for transmon qubit 1 ($V_{g_1}=\SI{0.75}{V}$) and while transmon qubit 2 is frequency-detuned with pinched-off nanowire Josephson junction ($V_{g_2}=\SI{-2}{V}$), we find a dispersive shift of $\chi=\SI{1.33}{MHz}$ and a cavity coupling rate of $\kappa=\SI{11.34}{MHz}$ and a setup-specific conversion factor $c=\SI{5.65}{mW^{-1}}$, see Fig.~\ref{fig:FigSphotonnumber}. For this configuration, the other system parameters are the resonator frequency $f_r=\SI{5.862}{GHz}$, the qubit frequency $f_{q_1}=\SI{4.827}{GHz}$ and the coherence time $T_1=\SI{1.3}{\mu s}$. Consequently, a signal generator input power of $P_{in}=\SI{-20}{dBm}$ corresponds to a cavity output power of $P_{out}=\SI{-138.09}{dBm}$ at a frequency of $f_r=\SI{5.862}{GHz}$. Since the transmon readout resonator is not resonant with the parametric amplifier, the SNR measurements presented in Fig.~\ref{fig:Fig4 Noise} were performed with a signal tone far detuned from the readout resonator frequency. Away from the passband of the readout resonator, we find that the transmission through the resonator is suppressed between 20 and \SI{25}{dB}. As such, the effective power level at the input of the parametric amplifier, when driven at the signal frequency around \SI{6.3}{GHz} with $P_{in}=\SI{-20}{dBm}$ at the signal generator, is expected to be in the range of \SI{-158}{dBm} to \SI{-163}{dBm}, which is consistent with the power level inferred from the line calibration and used in Fig.~\ref{fig:Fig4 Noise}b. Note that we were not able to accurately estimate the precise transmission at the signal frequency of the parametric amplifier due to an unexplained background signal in the transmission spectrum when measured through the readout resonator. Thus, we base the power level calibration in the main text solely on the line calibration data.

\section{Comparison of gate-tunable parametric amplifier with other existing implementations} \label{sec:comparison}

To put this work on gate-tunable and magnetic field compatible parametric amplifiers into perspective, we have prepared a table, see Tab. \ref{tab:amplifieroverview}, collecting many resonator based parametric amplifier parameters from a variety of publications addressing several design objectives such as a wide bandwidth, large saturation power, gate tunability, and magnetic field compatibility. From this table it is evident that our implementation is en par with other capacitively coupled SQUID, SQUID array and single junction implementations in terms of bandwidth and saturation power.

\begin{table*}[ht]
    \resizebox{1.8\columnwidth}{!}{
    \centering
    \begin{tabular}{|l|c|c|c|c|c|c|c|c|c|c|}
    \hline
        Paper & Year & Co. & Des. & WM & $f_r$ & $GBW$ & $P_{sat}$ & NLE & T & MF\\ \hline 
        & & & & & GHz & MHz & dBm & & & T\\ \hline \hline
        Castellanos \cite{Castellanos2007} & 2007 & C & DE & 4 & 7.8 & 3 & - & SQUID A & F & - \\
        Castellanos \cite{Castellanos2008} & 2008 & C & DE & 4 & 7 & 11 & - & SQUID A & F & - \\
        Yamamoto \cite{Yamamoto2008} & 2008 & C & CPW & 3 & 11 & 63 & -140 & SQUID & F & - \\
        Mutus \cite{Mutus2013} & 2013 & I & LE & 3,4 & 7 & - & -120 & SQUID & F & - \\
        Zhong \cite{Zhong2013} & 2013 & C & CPW & 3 & 5.6 & 15 & -136 & SQUID A & F & -  \\
        Zhou \cite{Zhou2014} & 2014 & C & LE & 3 & 6.0 & 60 & -123 & SQUID A & F & -  \\
        Mutus \cite{Mutus2014} & 2014 & K & LE & 3 & 6.6 & 3900 & -110 & SQUID & F & -\\
        Planat \cite{Planat2019} & 2019 & D & DE & 4 & 6.8 & 300 & -117 & SQUID A & F & - \\
        Sivak \cite{Sivak2019} & 2019 & C & DE & 3 & 7.2 & 250 & -102 & 3 SNAILs & F & - \\
        Sivak \cite{Sivak2020} & 2020 & C,I & DE & 3 & 12 & 110 & -108 & SNAILs & F & - \\
        Winkel \cite{Winkel2020} & 2020 & I & DE & 4 & 8.0 & 150 & -118 & 3 2 SQUID A & F & - \\
        Grebel \cite{Grebel2021} & 2021 & ST & LE & 3,4 & 5.3 & 3000 & -116 &  SQUID & F & -\\
        Lu \cite{Lu2022} & 2022 & IM & LE & 3 & 6.6 & 2500 & -114 &  SQUID & F & - \\
        White \cite{White2023} & 2022 & K & LE & 3 & 7 & 3000 & -95 & rf-SQUID A & F & -  \\
        Sarkar \cite{Sarkar2022} & 2022 & D & LE & 4 & 5.3 & 158 & -130 & graphene-JJ & G & -  \\
        Butseraen \cite{Butseraen2022} & 2022 & C & CPW & 4 & 6.2 & 33 & -123 & graphene-JJ & G & -  \\
        Ezenkova \cite{Ezenkova2022} & 2022 & ST & LE & 3 & 6.4 & 2100 & -100 & SNAILs & F & -  \\
        Parker \cite{Parker2022} & 2022 & BM & LE & 3 & 6.4 & 53 & - & NbTiN-KI & I & -  \\
        Qing \cite{Qing2023} & 2023 & K & LE & 3 & 7.1 & 3900 & -110 &  SQUID & F & - \\
        Xu \cite{Xu2023} & 2023 & C & LE & 4 & 7.5 & 59 & - &  NbN-KI & B & 0.5 \\
        Khalifa \cite{Khalifa2023} & 2023 & C & DE & 4 & 4.6 & - & - &  NbTiN-KI & B & 2 \\
        Phan \cite{Phan2022} & 2023 & C & CPW & 4 & 6.0 & 40 & -125 & 2DEG-JJ & G & 15m  \\
        \textit{this work} & 2023 & C & CPW & 4 & 6.1 & 30 & -120 & NW-KI & G & 0.5  \\
        Vaartjes \cite{Vaartjes2023} & 2023 & BM & LE & 3 & 6.2 & 17 & -86 & NbTiN-KI & I & 2 \\
        Frasca \cite{Frasca2023} & 2023 & ST & DE & 3 & 5.8 & 21 & -86 & NbN-KI & I & 6   \\ \hline 
    \end{tabular}
    }
    \caption{Parameter Overview of some resonator parametric amplifiers: Coupling (Co): (I) Inductive/galvanic, (K) Klopfenstein, (D) Direct, (C) Capacitive, (ST) Step transformer, (IM) 'fishbone' impedance matching, (BM) Bragg mirror; Design (Des): (LE) Lumped element, (DE) Distributed element, (CPW) Coplanar-Waveguide; Wavemixing (WM), Highest amplifier frequency $f_r$, Gain-Bandwidth product $GBW$, Input saturation power $P_{sat}$ for about \SI{20}{dB} weak signal gain, and nonlinear element NLE (single SQUIDS, SQUID arrays (SQUID A), asymmetric SQUIDS as SNAILS, rf-SQUID arrays, graphene based Josephson junctions, InAs 2DEG Josephson junctions, InAs nanowire kinetic inductances, and kinetic inductances from NbN and NbTiN); frequency tunability (T) achieved by flux (F), electrostatic gate (G), magnetic field (B) or current (I); Tested magnetic field compatibility up to the specified in-plane magnetic field.}
    \label{tab:amplifieroverview}
    
\end{table*}

\bibliography{nwpa_ref}

\end{document}